\documentclass[twocolumn,twoside,slac_two]{revtex4}
\usepackage{graphicx}
\usepackage{fancyhdr}
\usepackage{afterpage}
\newcommand{\rxj}{RX J1713.7-3946}
\newcommand{\gr}{$\gamma$-ray}

\newcommand{\geff}{\gamma_{\rm eff}}

\begin{document}

\title{Understanding hadronic $\gamma$-ray emission from supernova remnants}

\author{D. Caprioli}
\affiliation{INAF Osservatorio Astrofisico di Arcetri, L.go E. Fermi 5, 50125 Firenze, Italy}
\affiliation{Princeton University, 4 Ivy Ln, 08540 Princeton NJ, US}

\begin{abstract}
We aim to test the plausibility of a theoretical framework in which the $\gamma-$ray emission detected from supernova remnants is of hadronic origin, i.e., due to the decay of neutral pions produced in nuclear collisions involving relativistic nuclei. In particular, we investigate how the nature of the circumstellar medium affects the evolution of a remnant and of its $\gamma-$ray emission, stressing the role of magnetic field amplification in the prediction of expected particle spectra. A phenomenological scenario consistent with both the underlying Physics and the larger and larger amount of observational data provided by the present generation of $\gamma-$ray experiments is finally outlined and critically discussed.
\end{abstract}

\maketitle

\thispagestyle{fancy}

\section{$\gamma-$RAYS FROM GALACTIC SNRs}
Supernova Remnants (SNRs) have been regarded for many years as prominent sites for the acceleration of relativistic particles up to above $10^{6}$GeV \cite[see e.g.][for a review]{hillas05}. 
However, while the evidences of electron acceleration are clear-cut in the shape of synchrotron emission, a radiative evidence of the presence of non-thermal hadrons is much more elusive and has to be searched in the $\gamma-$rays resulting from the decay of neutral pions produced in nuclear collisions between cosmic rays (CRs) and the background plasma \citep[see e.g.][]{DAV94}.

\begin{figure}[tpb]
\centering
\includegraphics[width=0.48\textwidth]{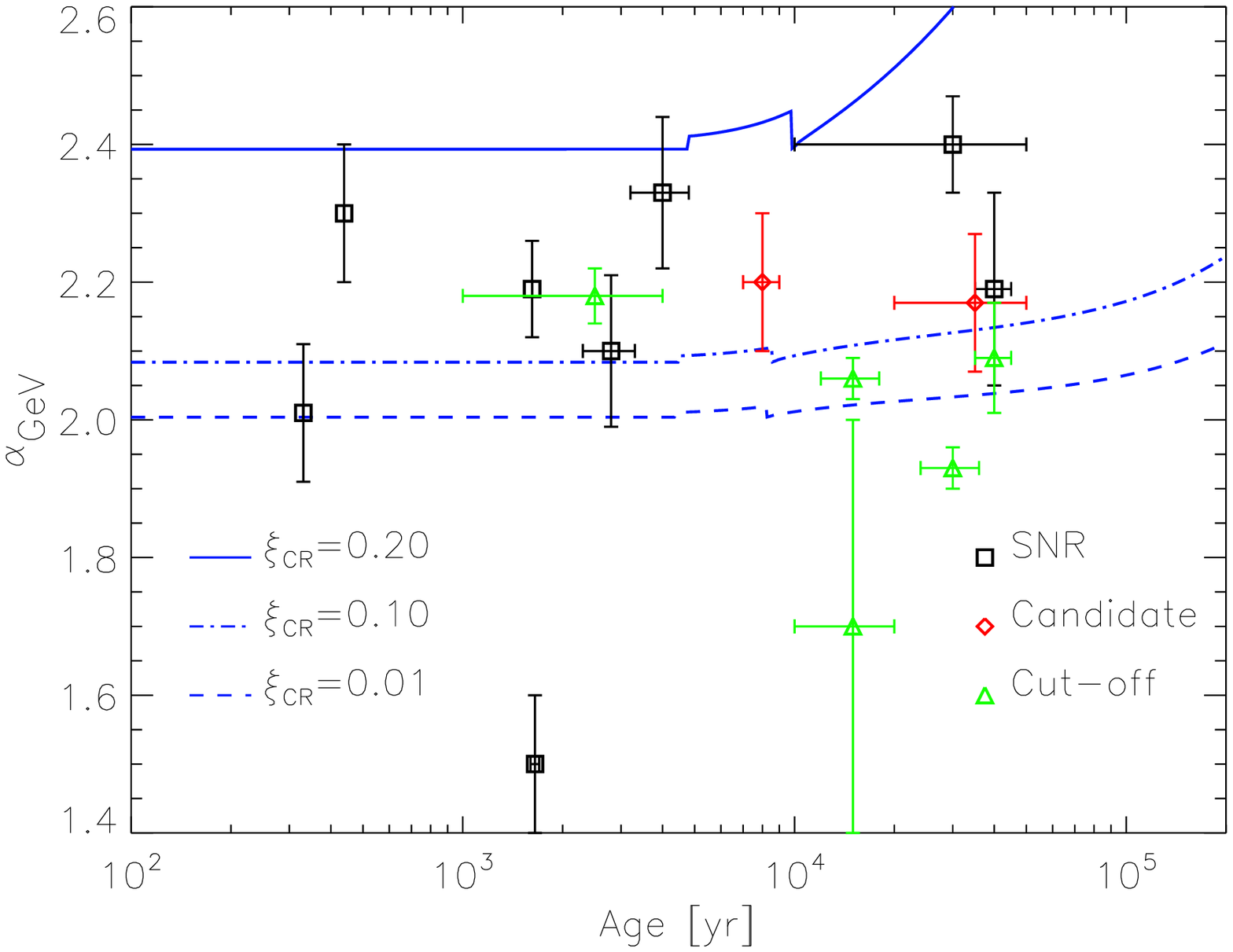}
\includegraphics[width=0.48\textwidth]{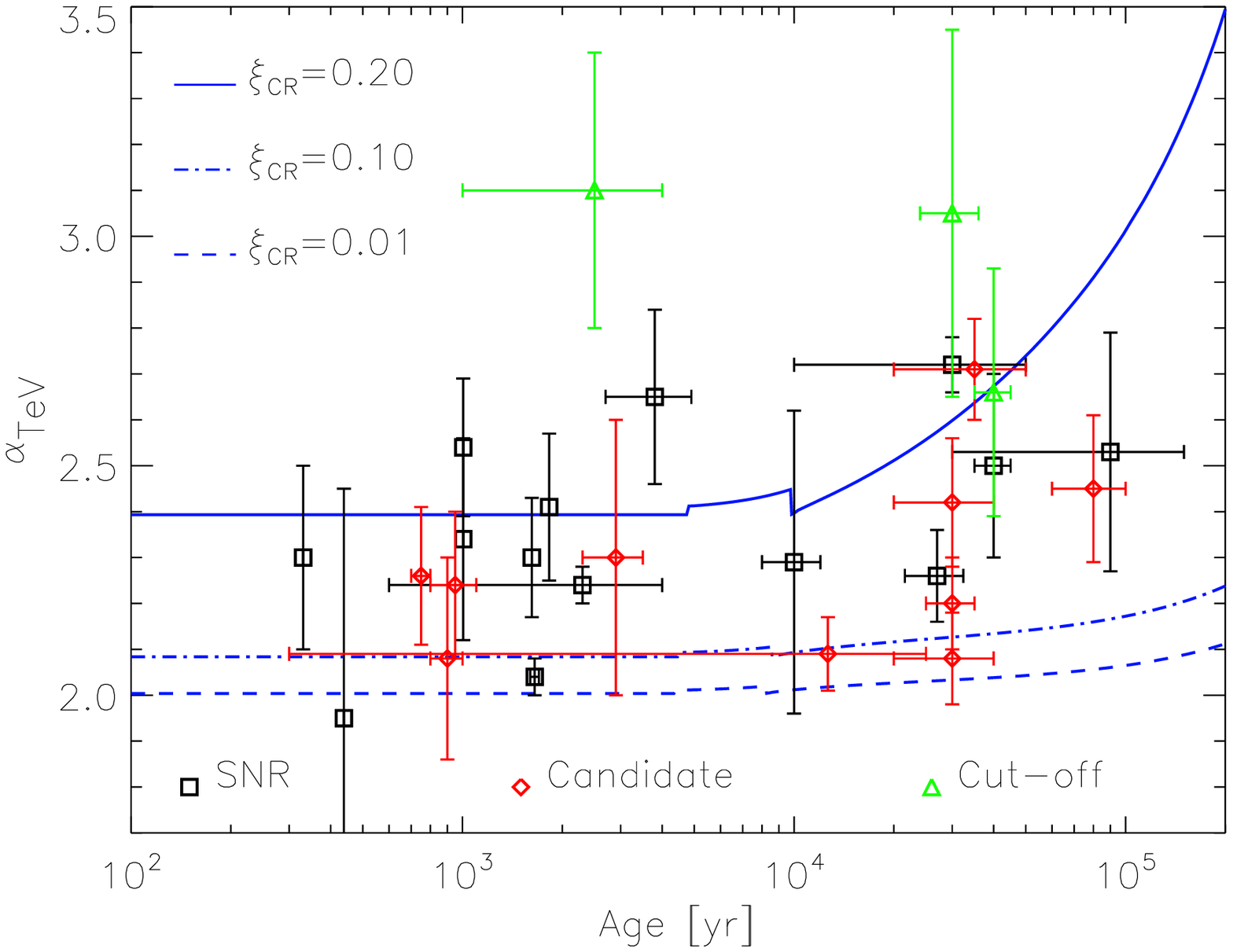}
\caption{Spectral slope inferred for $\gamma-$ray bright SNRs, as a function of their age. See table 1 in \protect{\cite{gamma}} for proper data references and \S\ref{sec:model} for the meaning of the different curves.} \label{slope}
\vspace{-0.5 cm}
\end{figure}

Unfortunately, the first observations of $\gamma-$ray bright SNRs have demonstrated themselves to be inconclusive for assessing the origin of the detected TeV emission, since also inverse Compton scattering of relativistic electrons on background photons may provide a signal in the same band.
A paradigmatic case in this respect is \rxj, the first SNR to be detected in $\gamma-$rays \citep{enomoto+02}, whose properties have been accounted for in both a hadronic or a leptonic scenario \citep[see e.g.][and references therein]{za10}.
Nevertheless, with the advent of the Fermi telescope, the detection of \rxj~also in the GeV band \citep{Fermi1713} eventually confirmed the leptonic nature of such an emission.

On the other hand, as showed by \cite{Tycho}, the first strong radiative evidence of hadronic emission from a Galactic SNR comes from another object very recently detected by VERITAS \citep{TychoVER} and Fermi-LAT \citep{TychoFermi}.
The reason why with broadband $\gamma-$ray observations it is possible to infer the origin of the emission relies on the fact that, while hadronic $\gamma-$rays show an energy spectrum parallel to the one of the parent nuclei, the inverse-Compton emission is instead flatter by a 0.5 in slope.
When looking at the TeV emission only, though, the cut-off in the electron distribution may show up and mimic the hadronic emission produced by a steep proton spectrum.
Nevertheless, this effect cannot extend to lower energies, too: provided that the primary spectrum goes like $\sim E^{-2}$ as predicted by first-order Fermi mechanism at strong shocks, any spectral slope in the GeV range larger than 1.5 may be regarded as a strong hint of hadron acceleration.

The main properties of $\gamma-$ray-bright Galactic SNRs are collected in table 1 of \cite{gamma}. 
In particular the inferred slope $\alpha$ of the photon spectrum, both in the GeV and TeV bands, is reported also in fig.~\ref{slope} as a function of the estimated age of the SNR.
Black squares correspond to confirmed SNRs, while red diamonds correspond to unidentified sources which have been proposed to be associated with SNRs. 
Green triangles, then, indicate SNRs whose Fermi-LAT spectrum exhibits a cut-off in the 1--20 GeV range: such a feature is hardly predictable both in leptonic and hadronic models for shell SNRs, and may be due to the superposition of different sources, e.g, pulsar-wind nebulae and/or molecular clouds (MCs).
The meaning of the curves in fig.~\ref{slope} will be explained in \S\ref{sec:model}.

The most striking evidence in fig.~\ref{slope} is that most of the detected SNRs show a photon spectral index larger than 2, and typically in the range $\alpha\sim$ 2.1--2.5.
As outlined above, while in the TeV band one may be looking at the cut-off in the particle spectrum, it is very unlikely to account for most of these GeV spectra in a leptonic scenario, since an electron spectrum as steep as $E^{-2.6}-E^{-3}$ would be required, at odds with what inferred, for instance, from radio measurements.  

Even more interestingly, such steep spectra is likely at odds also with what diffusive shock acceleration (DSA) at strong shocks predicts for accelerated nuclei.
In the test-particle limit the spectral slope depends only on the shock compression ratio, and is $\alpha=2$ for large Mach number shocks.
However, when the dynamical back-reaction of the accelerated particles is important, namely when the pressure in CRs becomes a sizable fraction of the bulk one, it is possible to show that the expected spectrum is no longer a power-law, being steeper (flatter) than $E^{-2}$ at low (high) energies \citep[see e.g.][for some reviews]{jones-ellison91,malkov-drury01}.
Since in this context ``low'' means below $\sim$ 1--10 GeV, the non-linear DSA theories which have been developed in the last three decades need to be revised under the light of the unprecedented wealth of data $\gamma-$ray satellites and Cherenkov telescopes are providing us with.
\vspace{-0.3cm}
\section{MODELS AND OBSERVATIONS}\label{sec:model}   
\vspace{-0.1cm}
In the last few years it has become clear that at SNR blast waves the magnetic field can be amplified by a factor of few tens with respect to the mean interstellar one (1--5$\mu$G) as a consequence of the super-Alfv\`enic streaming of the accelerated particles \citep[see e.g.][]{bell78a}. 
These large magnetic fields, beside scattering CRs more efficiently and having a dynamical role in the shock dynamics \citep{jumpl}, may as well significantly affect the shape of the spectrum of the accelerated particles.

CRs, in fact, only feel the compression ratio of the waves they scatter against: if Alfv\`en waves have a finite velocity $v_{w}$ with respect to the background plasma, the actual compression CRs experience is 
\begin{equation}\label{ratios}
r=\frac{u_{1}+v_{w,1}}{u_{2}+v_{w,2}},
\end{equation}
where $u$ is the fluid velocity and subscripts 1 and 2 refer to pre- and post-shock quantities.
Upstream, self-generated Alfv\'en waves travel in direction opposite to the CR pressure gradient ($v_{w,1}=-v_{A,1}=-B_{1}/\sqrt{4\pi \rho_{1}}$), while efficient isotropization downstream implies $v_{w,2}=0$.
Moreover, when CR acceleration is efficient and the instability saturates, the Alfv\`enic Mach number ahead of the shock is calculated to be (see eq.~42 of \cite{jumpkin}):
\begin{equation}\label{eq:Ma1}
M_{A,1}=\frac{2}{\xi_{cr}}\frac{(1-\xi_{cr})}{2-\xi_{cr}}^{5/2},
\end{equation}
where $\xi_{cr}$ is the CR acceleration efficiency, expressed here as the pressure in accelerated particles at the shock over the bulk pressure.
Following the simple algebra sketched in \S2 of \cite{gamma}, the effective shock compression ratio at the presence of efficient CR acceleration and magnetic field amplification reads:
\begin{equation}\label{eq:r}
r=\frac{\geff+1}{\geff-1+2/M_{s}^{2}}\left[1-\frac{\xi_{cr}(2-\xi_{cr})}{2(1-\xi_{cr})^{5/2}}\right],
\end{equation}
where $M_{s}$ is the sonic Mach number of the shock and $\geff=\frac{1}{3}\frac{5+3\xi_{cr}}{1+\xi_{cr}}$ is the effective adiabatic index of the fluid composed by CRs+thermal plasma.
\begin{figure}
\centering
	{\includegraphics[width=0.5\textwidth]{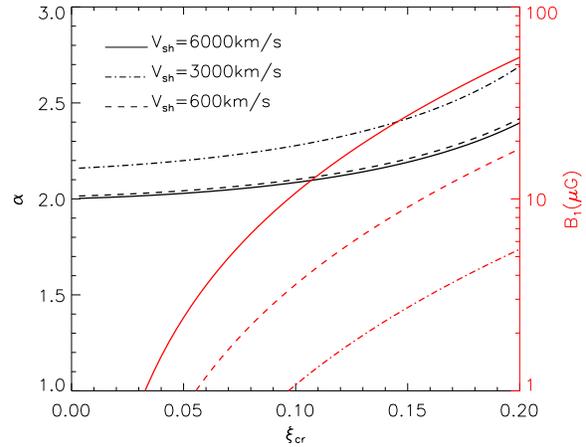}}
	\caption{Slope of the energy spectrum of accelerated particles (black lines, left axis) and self-amplified magnetic field upstream of the shock (red lines, right axis) as a function of the CR efficiency (see eqs.~\ref{eq:Ma1}--\ref{eq:alfa}).}
	\label{eff}
\vspace{-0.5 cm}
\end{figure}

The predicted slope for the spectrum of the accelerated particles is therefore simply given by
\begin{equation}\label{eq:alfa}
\alpha=\frac{r+2}{r-1}.
\end{equation}

In fig.~\ref{eff} the upstream self-generated magnetic field for different Mach numbers (temperature $T=10^{5}$K, number density $n=0.1$/cm$^{3}$, velocity as in the legend) is shown on the right axis, as a function of $\xi_{cr}$.
The corresponding spectral indexes expected for accelerated particles are shown on the left axis, and are larger and larger with increasing $\xi_{cr}$: this fact completely reverses the usual prediction that the more efficient the CR acceleration, the flatter the spectrum of the accelerated particles.

The simple calculation above has been worked out within the so called two-fluid approach, in which the CRs are modeled as a relativistic fluid, and has therefore to be checked against a kinetic modeling of particle acceleration in which the momentum dependence of the spectral slope is self-consistently accounted for. 
However, preliminary results outlined in \S 5.1 of \cite{jumpkin} support the scenario in which the whole particle spectrum may steepen significantly when the scattering centers have a non-negligible velocity with respect to the fluid.   
\vspace{-0.3cm}
\section{EVOLUTION OF THE $\gamma$-RAY EMISSION}
\vspace{-0.1cm}
In order to illustrate the prediction of the phenomenological model above, we consider a SNR produced by the explosion of a very massive star which underwent different stages of mass ejection, and in particular a slow and dense red-giant wind and a fast and hot Wolf-Rayet wind \citep[see e.g.][and references therein]{pz05}.

About 80\% of the Galactic SNRs are expected to be due to the core-collapse of massive progenitors \citep{hl05}, therefore the circumstellar medium their shocks propagate into should be ``polluted'' by winds launched in the pre-SN stages.
The case considered here is not meant to describe either a given or the most common object: it has to be regarded only as a benchmark case for investigating the effects of non-trivial environments on the expected hadronic $\gamma-$ray luminosity of SNRs, actually generalizing the pioneering work of \cite{DAV94}. 

\begin{figure}
		\includegraphics[width=0.5\textwidth]{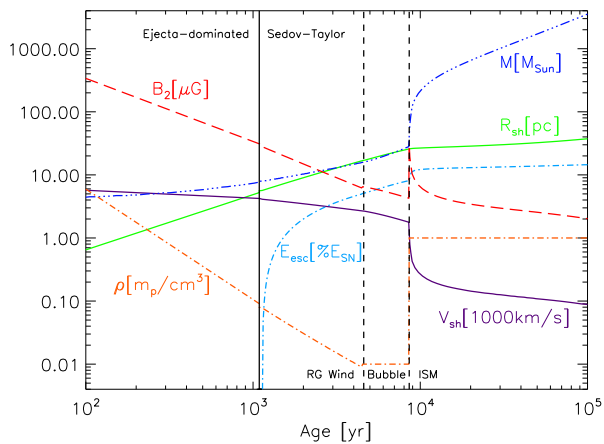}
	\caption{Time evolution of relevant physical quantities for $\xi_{cr}=0.1$. 
	The solid line around 1000 yr indicates the transition between ejecta-dominated and Sedov-Taylor stages, while the vertical dashed lines, from left to right, mark the boundaries of wind zone, hot bubble and ISM, as in the labels (see \S 3 of \cite{gamma} for details).}
	\label{Hydro}
\end{figure}

The details of the semi-analytical model (usually referred to as \emph{thin-shell approximation}) adopted to follow the shock propagation are widely discussed in \S 3 of \cite{gamma}.
What is relevant for our purposes is summarized in fig.~\ref{Hydro}: the shock (radius $R_{sh}$ and velocity $V_{sh}$) propagates first in the dense red-giant wind (density profile $\rho\propto r^{-2}$) and, for the parameters chosen, enters the Sedov-Taylor stage around 1000 yr. 
During the Sedov stage the most energetic CRs are expected to escape the system \citep{escape} carrying away a sizable amount of energy $E_{esc}$.
This energy loss is consistently accounted for in calculating the blast wave evolution.
At about 4000 yr the shock enters the hot and rarefied bubble excavated by the fast ($\sim 2000$km/s) Wolf-Rayet wind and finally, at about 10000 yr, the shock breaks into the unperturbed interstellar medium (ISM, $n=1/$cm$^{3}$, $T=10^{4}$K).     

When applying the formalism of \S\ref{sec:model}, the expected spectral slope as a function of the SNR age is plotted in fig.~\ref{slope} for three different CR acceleration efficiencies $\xi_{cr}=0.01$ (almost test-particle), 0.1 and 0.2, as in the legend.
There are two very interesting points arising from this kind of analysis:
\begin{itemize}
\item $\alpha$ is a rather strong function of the CR efficiency, but a moderate $\xi_{cr}$ between 0.1 and 0.2 may account for basically all the inferred slopes; 
\item for large Mach numbers the CR spectral slope is expected to be a function of the acceleration efficiency alone, and to increase when the shock velocity drops because of its encounter with the dense ISM (after 10--20$\times 10^{4}$ yr).  
\end{itemize}
Clearly these qualitative results have to be checked against proper non-linear, kinetic, models for DSA, but they indeed represent a possible way to reconcile our current understanding of particle acceleration in SNRs, magnetic field amplification due to CR-induced streaming instability and current $\gamma-$ray detections. 
The magnetic fields needed are in fact perfectly consistent with the one inferred by X-ray measurements.

\subsection{The role of the circumstellar medium}
\begin{figure}
\centering
	{\includegraphics[width=0.5\textwidth]{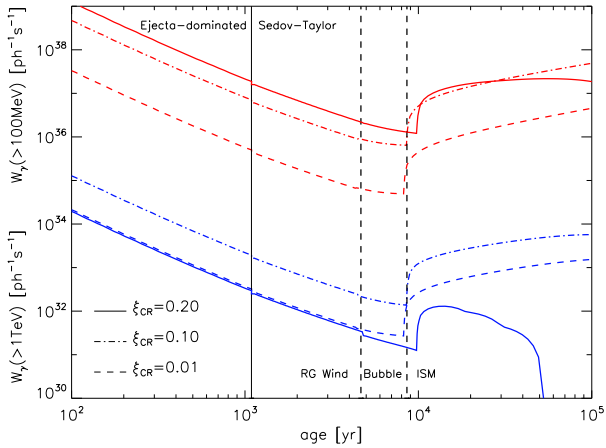}}
	\caption{Time evolution of the \gr~luminosity, in the GeV (upper lines) and in the TeV band (lower lines). 
	Different lines correspond to different CR acceleration efficiency, as in the legend.
	Vertical lines illustrate the evolutive stages for the case $\xi_{cr}=0.1$, as in figure \ref{Hydro}.}	\label{power}
	\vspace{-0.5 cm}
\end{figure}

The simple model presented here also allows us to track the expected $\gamma-$ray luminosity of the SNR during its evolution, and in particular to check if the (rapidly growing) statistics of bright SNRs might tell us something about the nature of the circumstellar medium they are propagating into, or even give us a hint of when a SNR is expected to be most luminous. 
The predicted $\gamma-$ray flux above 100 MeV and 1 TeV is shown in fig.~\ref{power} as a function of the SNR age.

A natural effect of the propagation through a rarefied bubble ($n< 0.01/$cm$^{3}$) is in fact a net decrease in the expected photon flux.
Provided that the statistics might not be large enough yet, and also considered that it is quite hard to estimate correctly the age of middle-age/old SNRs, the data in fig.~\ref{slope} are consistent with a bimodal distribution, with two populations separated by a gap in ages around 5--10$\times10^{3}$ yr.  
Further studies may confirm or disprove such a hint, but indeed the present and future generation of $\gamma-$ray telescopes could help us in understanding if CR acceleration occurs mainly in standard ISM or in superbubbles rich in Wolf-Rayet-like stars. 

\subsection{Caveats and further comments}
The scenario depicted above may be significantly affected by the presence of MCs interacting with, or simply close to, the SNR shock.
Many of the listed sources are in fact known to be associated with MCs \citep{jiang+10,cs10} which, on one hand, provide a large amount of targets enhancing the $\gamma-$ray luminosity but, on the other hand, represent a serious challenge for acceleration theories because of the issues related to shock propagation and particle diffusion in partially-ionized media.
In addition, some of these sources (W28N, IC443, W51C and W49B, represented by green triangles in fig.~\ref{slope}) exhibit a cut-off around 1--20 GeV and a very steep spectrum in the TeV band.
While their position in fig.~\ref{slope} is not extremely peculiar, it is likely that this class of sources is different from classical shell SNRs, therefore requiring a dedicated description.

The model above, while retaining some key ingredients of CR-modified shocks, does not include any information about the maximum energy achieved by particles at a given SNR age. 
The expected decrease of this quantity with the shock velocity would eventually lead the $\gamma-$ray emission to fade at late evolutive stages, unless some high-energy particles remain trapped in the downstream. 
Such an effect, however, is very hard to quantify and is tightly related to the interesting problem of how accelerated particles leave their sources and become CRs \citep{crspectrum}.
 
When focusing on a given object rather than trying a statistical approach to many Galactic remnants, it is indeed possible to implement more self-consistently the idea that magnetic field amplification may play a crucial role in explaining hadronic emission from SNRs.
A nice example has been put forward by \citep{Tycho}, who clearly show that Tycho's shock channels a sizable fraction ($\sim 10\%$) of its kinetic energy into hadrons, accelerated up to at least $5\times 10^{5}$GeV in magnetic fields as large as $\sim 300\mu$G.

\vspace{-0.4cm}

\bibliography{bibFS}

\begin{thebibliography}{20}
\expandafter\ifx\csname natexlab\endcsname\relax\def\natexlab#1{#1}\fi
\expandafter\ifx\csname bibnamefont\endcsname\relax
  \def\bibnamefont#1{#1}\fi
\expandafter\ifx\csname bibfnamefont\endcsname\relax
  \def\bibfnamefont#1{#1}\fi
\expandafter\ifx\csname citenamefont\endcsname\relax
  \def\citenamefont#1{#1}\fi
\expandafter\ifx\csname url\endcsname\relax
  \def\url#1{\texttt{#1}}\fi
\expandafter\ifx\csname urlprefix\endcsname\relax\def\urlprefix{URL }\fi
\providecommand{\bibinfo}[2]{#2}
\providecommand{\eprint}[2][]{\url{#2}}

\bibitem[{\citenamefont{{Hillas}}(2005)}]{hillas05}
\bibinfo{author}{\bibfnamefont{A.~M.} \bibnamefont{{Hillas}}},
  \bibinfo{journal}{Journal of Physics G Nuclear Physics}
  \textbf{\bibinfo{volume}{31}}, \bibinfo{pages}{95} (\bibinfo{year}{2005}).

\bibitem[{\citenamefont{{O'C. Drury} et~al.}(1994)\citenamefont{{O'C. Drury},
  {Aharonian}, and {V{\"o}lk}}}]{DAV94}
\bibinfo{author}{\bibfnamefont{L.}~\bibnamefont{{O'C. Drury}}},
  \bibinfo{author}{\bibfnamefont{F.}~\bibnamefont{{Aharonian}}},
  \bibnamefont{and} \bibinfo{author}{\bibfnamefont{H.~J.}
  \bibnamefont{{V{\"o}lk}}}, \bibinfo{journal}{\aap}
  \textbf{\bibinfo{volume}{287}}, \bibinfo{pages}{959} (\bibinfo{year}{1994}),
  \eprint{astro-ph/9305037}.

\bibitem[{\citenamefont{{Caprioli}}(2011)}]{gamma}
\bibinfo{author}{\bibfnamefont{D.}~\bibnamefont{{Caprioli}}},
  \bibinfo{journal}{\jcap} \textbf{\bibinfo{volume}{5}}, \bibinfo{pages}{26}
  (\bibinfo{year}{2011}), \eprint{1103.2624}.

\bibitem[{\citenamefont{{Enomoto et al.}}(2002)}]{enomoto+02}
\bibinfo{author}{\bibfnamefont{R.}~\bibnamefont{{Enomoto et al.}}},
  \bibinfo{journal}{Nature} \textbf{\bibinfo{volume}{416}},
  \bibinfo{pages}{823} (\bibinfo{year}{2002}).

\bibitem[{\citenamefont{{Zirakashvili} and {Aharonian}}(2010)}]{za10}
\bibinfo{author}{\bibfnamefont{V.~N.} \bibnamefont{{Zirakashvili}}}
  \bibnamefont{and} \bibinfo{author}{\bibfnamefont{F.~A.}
  \bibnamefont{{Aharonian}}}, \bibinfo{journal}{ApJ}
  \textbf{\bibinfo{volume}{708}}, \bibinfo{pages}{965} (\bibinfo{year}{2010}),
  \eprint{0909.2285}.

\bibitem[{\citenamefont{{Abdo} and {Fermi LAT
  Collaboration}}(2011)}]{Fermi1713}
\bibinfo{author}{\bibfnamefont{A.~A.} \bibnamefont{{Abdo}}} \bibnamefont{and}
  \bibinfo{author}{\bibnamefont{{Fermi LAT Collaboration}}},
  \bibinfo{journal}{ArXiv e-prints}  (\bibinfo{year}{2011}),
  \eprint{1103.5727}.

\bibitem[{\citenamefont{{Morlino} and {Caprioli}}(2011)}]{Tycho}
\bibinfo{author}{\bibfnamefont{G.}~\bibnamefont{{Morlino}}} \bibnamefont{and}
  \bibinfo{author}{\bibfnamefont{D.}~\bibnamefont{{Caprioli}}},
  \bibinfo{journal}{ArXiv e-prints}  (\bibinfo{year}{2011}),
  \eprint{1105.6342}.

\bibitem[{\citenamefont{{Acciari et al.}}(2011)}]{TychoVER}
\bibinfo{author}{\bibfnamefont{V.~A.} \bibnamefont{{Acciari et al.}}},
  \bibinfo{journal}{ArXiv e-prints}  (\bibinfo{year}{2011}),
  \eprint{1102.3871}.

\bibitem[{\citenamefont{{Giordano} et~al.}(2011)\citenamefont{{Giordano},
  {Naumann-Godo}, {Ballet}, {Bechtol}, {Funk}, {Lande}, {Mazziotta},
  {Rain{\`o}}, {Tanaka}, {Tibolla} et~al.}}]{TychoFermi}
\bibinfo{author}{\bibfnamefont{F.}~\bibnamefont{{Giordano}}},
  \bibinfo{author}{\bibfnamefont{M.}~\bibnamefont{{Naumann-Godo}}},
  \bibinfo{author}{\bibfnamefont{J.}~\bibnamefont{{Ballet}}},
  \bibinfo{author}{\bibfnamefont{K.}~\bibnamefont{{Bechtol}}},
  \bibinfo{author}{\bibfnamefont{S.}~\bibnamefont{{Funk}}},
  \bibinfo{author}{\bibfnamefont{J.}~\bibnamefont{{Lande}}},
  \bibinfo{author}{\bibfnamefont{M.~N.} \bibnamefont{{Mazziotta}}},
  \bibinfo{author}{\bibfnamefont{S.}~\bibnamefont{{Rain{\`o}}}},
  \bibinfo{author}{\bibfnamefont{T.}~\bibnamefont{{Tanaka}}},
  \bibinfo{author}{\bibfnamefont{O.}~\bibnamefont{{Tibolla}}},
  \bibnamefont{et~al.}, \bibinfo{journal}{ArXiv e-prints}
  (\bibinfo{year}{2011}), \eprint{1108.0265}.

\bibitem[{\citenamefont{{Malkov} and {O'C. Drury}}(2001)}]{malkov-drury01}
\bibinfo{author}{\bibfnamefont{M.~A.} \bibnamefont{{Malkov}}} \bibnamefont{and}
  \bibinfo{author}{\bibfnamefont{L.}~\bibnamefont{{O'C. Drury}}},
  \bibinfo{journal}{Reports of Progress in Physics}
  \textbf{\bibinfo{volume}{64}}, \bibinfo{pages}{429} (\bibinfo{year}{2001}).

\bibitem[{\citenamefont{{Jones} and {Ellison}}(1991)}]{jones-ellison91}
\bibinfo{author}{\bibfnamefont{F.~C.} \bibnamefont{{Jones}}} \bibnamefont{and}
  \bibinfo{author}{\bibfnamefont{D.~C.} \bibnamefont{{Ellison}}},
  \bibinfo{journal}{Space Science Reviews} \textbf{\bibinfo{volume}{58}},
  \bibinfo{pages}{259} (\bibinfo{year}{1991}).

\bibitem[{\citenamefont{{Bell}}(1978)}]{bell78a}
\bibinfo{author}{\bibfnamefont{A.~R.} \bibnamefont{{Bell}}},
  \bibinfo{journal}{MNRAS} \textbf{\bibinfo{volume}{182}}, \bibinfo{pages}{147}
  (\bibinfo{year}{1978}).

\bibitem[{\citenamefont{{Caprioli} et~al.}(2008)\citenamefont{{Caprioli},
  {Blasi}, {Amato}, and {Vietri}}}]{jumpl}
\bibinfo{author}{\bibfnamefont{D.}~\bibnamefont{{Caprioli}}},
  \bibinfo{author}{\bibfnamefont{P.}~\bibnamefont{{Blasi}}},
  \bibinfo{author}{\bibfnamefont{E.}~\bibnamefont{{Amato}}}, \bibnamefont{and}
  \bibinfo{author}{\bibfnamefont{M.}~\bibnamefont{{Vietri}}},
  \bibinfo{journal}{\apjl} \textbf{\bibinfo{volume}{679}},
  \bibinfo{pages}{L139} (\bibinfo{year}{2008}), \eprint{0804.2884}.

\bibitem[{\citenamefont{{Caprioli}
  et~al.}(2009{\natexlab{a}})\citenamefont{{Caprioli}, {Blasi}, {Amato}, and
  {Vietri}}}]{jumpkin}
\bibinfo{author}{\bibfnamefont{D.}~\bibnamefont{{Caprioli}}},
  \bibinfo{author}{\bibfnamefont{P.}~\bibnamefont{{Blasi}}},
  \bibinfo{author}{\bibfnamefont{E.}~\bibnamefont{{Amato}}}, \bibnamefont{and}
  \bibinfo{author}{\bibfnamefont{M.}~\bibnamefont{{Vietri}}},
  \bibinfo{journal}{\mnras} \textbf{\bibinfo{volume}{395}},
  \bibinfo{pages}{895} (\bibinfo{year}{2009}{\natexlab{a}}),
  \eprint{0807.4261}.

\bibitem[{\citenamefont{{Ptuskin} and {Zirakashvili}}(2005)}]{pz05}
\bibinfo{author}{\bibfnamefont{V.~S.} \bibnamefont{{Ptuskin}}}
  \bibnamefont{and} \bibinfo{author}{\bibfnamefont{V.~N.}
  \bibnamefont{{Zirakashvili}}}, \bibinfo{journal}{\aap}
  \textbf{\bibinfo{volume}{429}}, \bibinfo{pages}{755} (\bibinfo{year}{2005}),
  \eprint{astro-ph/0408025}.

\bibitem[{\citenamefont{{Higdon} and {Lingenfelter}}(2005)}]{hl05}
\bibinfo{author}{\bibfnamefont{J.~C.} \bibnamefont{{Higdon}}} \bibnamefont{and}
  \bibinfo{author}{\bibfnamefont{R.~E.} \bibnamefont{{Lingenfelter}}},
  \bibinfo{journal}{\apj} \textbf{\bibinfo{volume}{628}}, \bibinfo{pages}{738}
  (\bibinfo{year}{2005}).

\bibitem[{\citenamefont{{Caprioli}
  et~al.}(2009{\natexlab{b}})\citenamefont{{Caprioli}, {Blasi}, and
  {Amato}}}]{escape}
\bibinfo{author}{\bibfnamefont{D.}~\bibnamefont{{Caprioli}}},
  \bibinfo{author}{\bibfnamefont{P.}~\bibnamefont{{Blasi}}}, \bibnamefont{and}
  \bibinfo{author}{\bibfnamefont{E.}~\bibnamefont{{Amato}}},
  \bibinfo{journal}{\mnras} \textbf{\bibinfo{volume}{396}},
  \bibinfo{pages}{2065} (\bibinfo{year}{2009}{\natexlab{b}}),
  \eprint{0807.4259}.

\bibitem[{\citenamefont{{Jiang} et~al.}(2010)\citenamefont{{Jiang}, {Chen},
  {Wang}, {Su}, {Zhou}, {Safi-Harb}, and {DeLaney}}}]{jiang+10}
\bibinfo{author}{\bibfnamefont{B.}~\bibnamefont{{Jiang}}},
  \bibinfo{author}{\bibfnamefont{Y.}~\bibnamefont{{Chen}}},
  \bibinfo{author}{\bibfnamefont{J.}~\bibnamefont{{Wang}}},
  \bibinfo{author}{\bibfnamefont{Y.}~\bibnamefont{{Su}}},
  \bibinfo{author}{\bibfnamefont{X.}~\bibnamefont{{Zhou}}},
  \bibinfo{author}{\bibfnamefont{S.}~\bibnamefont{{Safi-Harb}}},
  \bibnamefont{and}
  \bibinfo{author}{\bibfnamefont{T.}~\bibnamefont{{DeLaney}}},
  \bibinfo{journal}{\apj} \textbf{\bibinfo{volume}{712}}, \bibinfo{pages}{1147}
  (\bibinfo{year}{2010}), \eprint{1001.2204}.

\bibitem[{\citenamefont{{Castro} and {Slane}}(2010)}]{cs10}
\bibinfo{author}{\bibfnamefont{D.}~\bibnamefont{{Castro}}} \bibnamefont{and}
  \bibinfo{author}{\bibfnamefont{P.}~\bibnamefont{{Slane}}},
  \bibinfo{journal}{\apj} \textbf{\bibinfo{volume}{717}}, \bibinfo{pages}{372}
  (\bibinfo{year}{2010}), \eprint{1002.2738}.

\bibitem[{\citenamefont{{Caprioli} et~al.}(2010)\citenamefont{{Caprioli},
  {Amato}, and {Blasi}}}]{crspectrum}
\bibinfo{author}{\bibfnamefont{D.}~\bibnamefont{{Caprioli}}},
  \bibinfo{author}{\bibfnamefont{E.}~\bibnamefont{{Amato}}}, \bibnamefont{and}
  \bibinfo{author}{\bibfnamefont{P.}~\bibnamefont{{Blasi}}},
  \bibinfo{journal}{APh} \textbf{\bibinfo{volume}{33}}, \bibinfo{pages}{160}
  (\bibinfo{year}{2010}), \eprint{0912.2964}.

\end{thebibliography}

\end{document}